\documentclass[fleqn,usenatbib]{mnras}

\usepackage{newtxtext,newtxmath}

\usepackage[T1]{fontenc}
\usepackage{ae,aecompl}
\usepackage{euscript}

\usepackage{graphicx}	
\usepackage{amsmath}	
\usepackage{amssymb}	
\makeatletter
\newcommand*{\rom}[1]{\expandafter\@slowromancap\romannumeral #1@}
\newcommand*{\Rom}[1]{\expandafter\@slowromancap\romannumeral #1@}
\makeatother

\title[SXP 91.1]{An X-ray and optical study of the outbursting behaviour of the SMC Be X-ray binary SXP 91.1}

\author[Monageng et al.]
  {I.M. Monageng$^{1}$\thanks{E-mail: itu@saao.ac.za}, 
  M.J. Coe$^{2}$,
  J.A. Kennea$^3$, 
  L.J. Townsend$^4$,
  D.A.H. Buckley$^1$,
  \newauthor
  V.A. McBride$^1$,
  A. Udalski$^5$, 
  P.A. Evans$^6$,
  P.D. Roche $^7$
\\
$^1$South African Astronomical Observatory, P.O Box 9, Observatory, 7935, Cape Town, South Africa\\
$^2$Physics \& Astronomy, University of Southampton, SO17 1BJ, UK\\
$^3$Department of Astronomy and Astrophysics, The Pennsylvania State University, University Park, PA 16802, USA\\
$^4$Department of Astronomy, University of Cape Town, Private Bag X3, Rondebosch 7701, South Africa\\
$^5$Warsaw University Observatory, Al. Ujazdowskie 4, 00-478 Warszawa, Poland\\
$^6$University of Leicester, X-ray and Observational Astronomy Research Group, Leicester Institute for Space and Earth Observation, \\
Department of Physics \& Astronomy, University Road, Leicester LE1 7RH, UK\\
$^7$School of Physics \& Astronomy, Cardiff University, The Parade, Cardiff, CF24 3AA\\
}

\date{Accepted XXX. Received YYY; in original form ZZZ}

\pubyear{2019}

\begin{document}
\label{firstpage}
\pagerange{\pageref{firstpage}--\pageref{lastpage}}
\maketitle

\begin{abstract}
In this paper we report on the optical and X-ray behaviour of the Be X-ray binary, SXP 91.1, during a recent type~\Rom{1} outburst. We monitored the outburst using the Neil Gehrels \textit{Swift} Observatory. These data were supported by optical data from the Southern African Large Telescope (SALT) and the Optical Gravitational Lensing Experiment (OGLE) to show the circumstellar disc activity. Matter from this disc accretes onto the neutron star, giving rise to the X-ray outburst as seen in the synchronous evolution of the optical and X-ray lightcurves. Using data taken with OGLE we show that the circumstellar disc has exhibited stable behaviour over two decades. A positive correlation is seen between the colour and magnitude from the OGLE and MACHO observations, which indicates that the disc is orientated at relatively low inclination angles. From the OGLE and \textit{Swift} data, we demonstrate that the system has shown relative phase offsets that have persisted for many years. The spin period derivative is seen to be at maximum spin-up at phases when the mass accretion rate is at maximum. We show that the neutron star in SXP 91.1 is an unusual member of its class in that it has had a consistent spin period derivative over many years, with the average spin-up rate being one of the highest for known SMC pulsars. The most recent measurements of the spin-up rate reveal higher values than the global trend, which is attributed to the recent mass accretion event leading to the current outburst..
\end{abstract}




\section{Introduction}

High mass X-ray binaries (HMXBs) comprise a massive early-type star (O or B spectral type) and a compact object (neutron star or black hole). By convention, they are divided up into two sub-groups on the basis of their luminosity class: supergiant X-ray binaries (luminosity class \Rom{1} and \Rom{2}) and Be X-ray binaries (\Rom{3}, \Rom{4} and \Rom{5}). In Be X-ray binaries (BeXBs) the massive companion has a geometrically thin Keplerian disc from which the compact object, primarily a neutron star (NS), accretes matter, resulting in X-ray outbursts. The X-ray outbursts come in two flavours: type \rom{1} (normal) and type \Rom{2} (giant). Type \Rom{1} outbursts ($L \leq 10^{37}$erg.s$^{-1}$) occur more frequently and are typically associated with the periastron passage of the NS, while type \Rom{2} outbursts ($L \geq 10^{37}$erg.s$^{-1}$; \citealt{1986ApJ...308..669S}) do not show any correlation with orbital phase and last longer. The physical mechanism by which type \Rom{2} outbursts occur is still not well-understood, with a number of models proposed - (see \citealt{2013PASJ...65...83M, 2014ApJ...790L..34M, 2017MNRAS.464..572M} for discussions). \\
The Small Magellanic Cloud (SMC) hosts a large population of HMXBs ($\sim$70; \citealt{2015MNRAS.452..969C}), which is comparable to that of the Milky Way (65; \citealt{2011MNRAS.413.1600R}) despite a large mass ratio of the two galaxies ($M_{MW}/M_{SMC} \sim $ 50). This is most likely attributed to a recent event of star formation due to an enhancement in the tidal force on the SMC from an interaction with the Large Magellanic Cloud \citep{2011MNRAS.413.2015D}.\\
The system discussed in this paper, SXP 91.1, was discovered in the \textit{Rossi X-ray Timing Explorer} (RXTE) survey when a period of 92$\pm$1.5 s was detected \citep{1997IAUC.6777....2M}. \cite{1998IAUC.6803....1C} later improved on this with a refined period 91.12$\pm$0.05~s. In a study of optical properties of SMC sources, \citet{1999MNRAS.309..421S} identified the optical counterpart with a strong H$\alpha$ emission line, typical of Be stars. \citet{2004AJ....127.3388S} reported an 88.25~day orbital period from \textit{MACHO} $V$ and $R$ data. This period was later refined by \citet{2012MNRAS.423.3663B} to $88.37\pm0.03$d using a larger \textit{OGLE} data base.\\
In this paper we report on the optical and X-ray behaviour of SXP 91.1 during its outburst in November 2018. We also review the system's historic behaviour.

\section{Observations}

\subsection{Swift}
\label{sec:Swift}

The source was monitored over the 0.3 - 10 keV range throughout the periastron passage by the  Neil Gehrels \textit{Swift} observatory \citep{2004ApJ...611.1005G}. The observations cover just over 50d, at an approximate cadence of once every 3d,  and typical exposure times used were $\sim1$ks. The XRT lightcurve was produced following the instructions described in the Swift data analysis guide\footnote{http://www.swift.ac.uk/analysis/xrt/}. The results from these observations are shown in Fig.~\ref{fig:OGLE_EW_Scubed}. At its peak brightness, and assuming the source is in the SMC (62~kpc; \citealt{2016ApJ...816...49S}), a luminosity of $9 \times 10^{35}$erg.s$^{-1}$ was measured from SXP 91.1. Further Swift/XRT data on this source covering the period 2016 - 2018 were obtained through the S-CUBE project \citep{2018ApJ...868...47K} and are also included in this paper.

\subsection{OGLE \& MACHO}
\label{sec:OGLE} 

The OGLE project \citep{1997AcA....47..319U, 2015AcA....65....1U}  provides long term $I$-band photometry with a typical cadence of 1-3 days. The optical counterpart of the source, the Be star [M2002] SMC 18187, has been observed for over 17 years in the $I$-band. The source is identified as SMC219.21.21951 in OGLE IV \citep{2015AcA....65....1U} and SMC102.1.32 in OGLE III \citep{2003AcA....53..291U}. The cadence of the observations was increased to daily measurements around the time of the outburst discussed in this paper, and the lightcurve of the $I$-band magnitudes is shown in Fig.~\ref{fig:OGLE_EW_Scubed}. 

The MAssive Compact Halo Objects project (MA-
CHO) conducted a survey producing regular photometric measurements
of several million Magellanic Cloud and Galactic bulge stars
\citep{1993ASPC...43..291A}. The data for SXP91.1 (with identifier 208.16034.5 in the MACHO catalogue) cover $\sim$6 years starting in 1992 and is available on the MACHO website\footnote{http://macho.nci.org.au/}. We use data taken in the red ($\mathcal{R}$; 6300-7600~\AA) and blue ($\mathcal{V}$; 4500-6300~\AA) passbands.

\subsection{SALT}
\label{sec:SALT}
The optical counterpart of SXP 91.1 was observed with the Southern African Large Telescope (SALT; \citealt{2006SPIE.6267E..0ZB}) using the Robert Stobie Spectrograph (RSS; \citealt{2003SPIE.4841.1463B,2003SPIE.4841.1634K}) and the High Resolution Spectrograph (HRS; \citealt{2010SPIE.7735E..4FB,2012SPIE.8446E..0AB,2014SPIE.9147E..6TC}). The RSS observations were done between 03 September 2017 and 29 November 2018 with grating PG1800 covering wavelength range $5985-7200$~\AA\AA~ with a resolution of 1.9~\AA. Single exposure times of 300~s were used for all the RSS observations. The SALT pipeline \citep{2012ascl.soft07010C} was used to perform the primary reductions (which include overscan correction, bias subtraction, gain correction and amplifier cross-talk correction). Subsequent steps of the reductions (identifying arc lines, subtraction of the background and extraction of the 1D spectra) were carried out using various tasks in \textsc{iraf}\footnote{Image Reduction and Analysis Facility: iraf.noao.edu}.\\
The HRS observations (18 November 2018, 24 November 2018, 02 December 2018) were done in low resolution mode (R$\sim 14 000$) with single exposure times of 1800~s covering a wavelength range of $\sim 5500-8800$~\AA\AA~ with a resolution of $\sim$0.4~\AA. The primary reductions of these were done with the SALT pipeline \citep{2015ascl.soft11005C}. The remainder of the reduction steps (background subtraction, identification of arc lines, blaze function removal and merging of the orders) were performed with the \textsc{midas feros} \citep{1999ASPC..188..331S} and \textsc{echelle} \citep{1992ESOC...41..177B} packages. \citet{2016MNRAS.459.3068K} describe the procedure in detail.

\subsection{NuStar}
\label{sec:NuStar}

Observations of the field which includes SXP 91.1 were carried out by the NuSTAR observatory  \citep{2013ApJ...770..103H} during November 2018 whilst studying the bright nearby X-ray transient SXP 4.78 \citep{2019MNRAS.485.4617M}. Data from the Focal Plane Module A (FPMA) and Focal Plane Module B (FPMB) instruments were used covering the energy range 3 - 79 keV. The result from three separate observations of SXP 91.1 over a period of 12 days are presented in Table~\ref{tab:period_measurements}.

\section{Results}
\subsection{Photometric and X-ray variability}
\begin{figure}
	\includegraphics[width=\columnwidth]{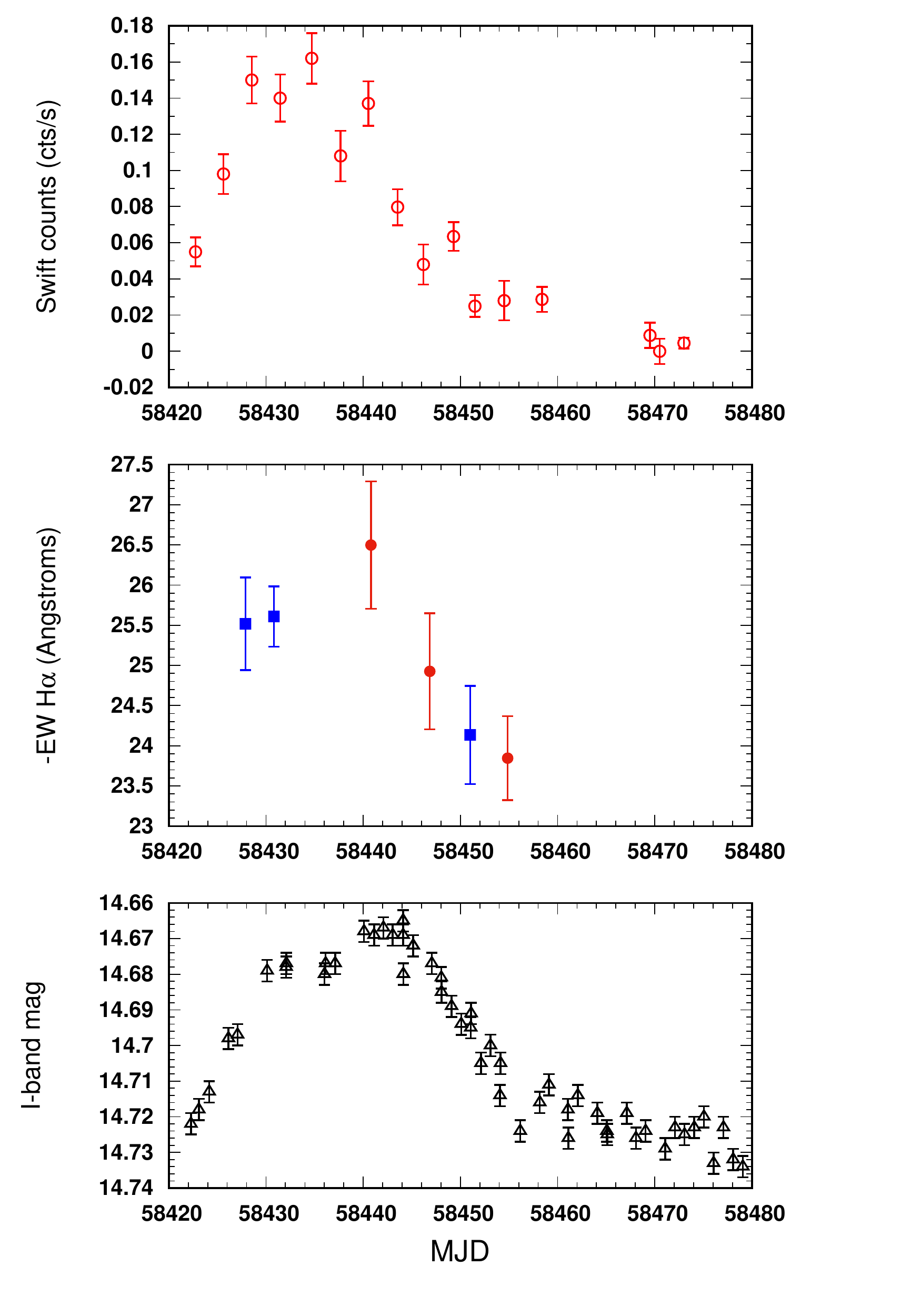}
    \caption{The Swift X-ray (top panel), H$\alpha$ EW (middle panel; blue squares are RSS observations and red circles are from HRS) and OGLE $I$-band measurements (bottom panel, black triangles) during the current outburst.} 
    \label{fig:OGLE_EW_Scubed}
\end{figure}

\begin{figure}
	\includegraphics[width=\columnwidth]{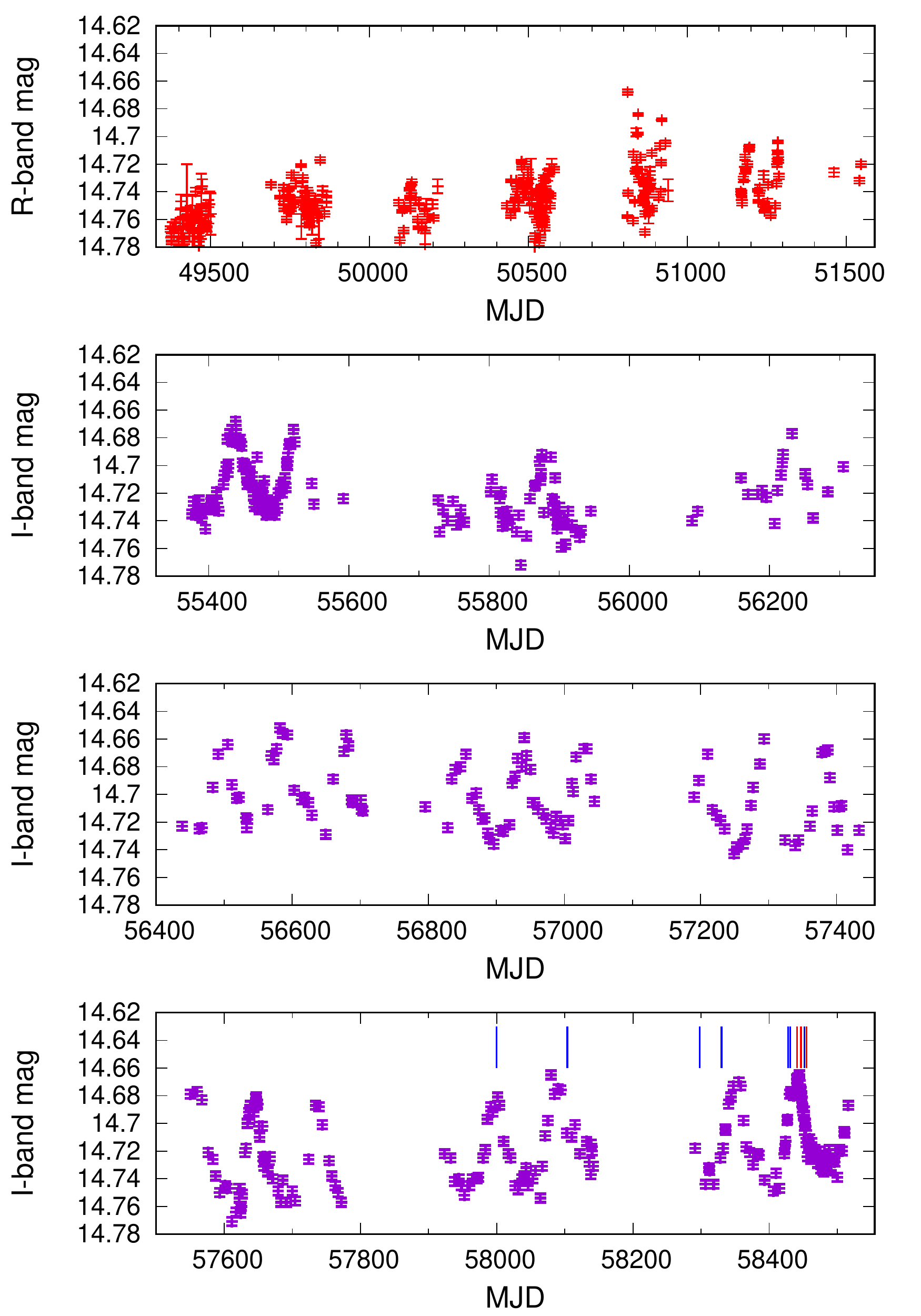}
    \caption{Long-term MACHO $\mathcal{R}$-band (red symbols) and OGLE $I$-band (purple symbols) lightcurve showing historic behaviour of SXP 91.1. The markers indicate epochs of SALT observations with RSS (blue symbols) and HRS (red symbols).}
    \label{fig:OGLE_multiplot}
\end{figure}

Fig.~\ref{fig:OGLE_EW_Scubed} shows the OGLE $I$-band, H$\alpha$ equivalent width (EW) and Swift X-ray measurements during the current outburst. From the figure, the intensities from the lightcurves follow each other as they rise and decline. It is well-known that the optical photometric variability in BeXRBs is attributed to changes in the circumstellar disc (e.g. \citealt{2011MNRAS.413.1600R}). In a simple picture, this suggests that the disc grew large enough in size for matter to be accreted by the NS resulting in the enhanced X-ray activity which peaks while the disc is still growing, before shrinking. \\ 
The long-term OGLE $I$-band lightcurve (Fig.~\ref{fig:OGLE_multiplot}) shows stable behaviour, with peaks separated by the orbital period with similar amplitudes ($\lesssim$0.1 mag) over a time period of $\sim$9 years. This is an indicator of disc truncation by the NS, as it sets a limit to the extent to which the disc can grow. \\
The correlation between the increase in X-ray activity and optical flux, as seen in the current outburst, appears to be a consistent historic feature of the system (see section~\ref{sec:longterm}).  \\

\subsection{Colour-index variability}
\begin{figure}
	\includegraphics[width=\columnwidth]{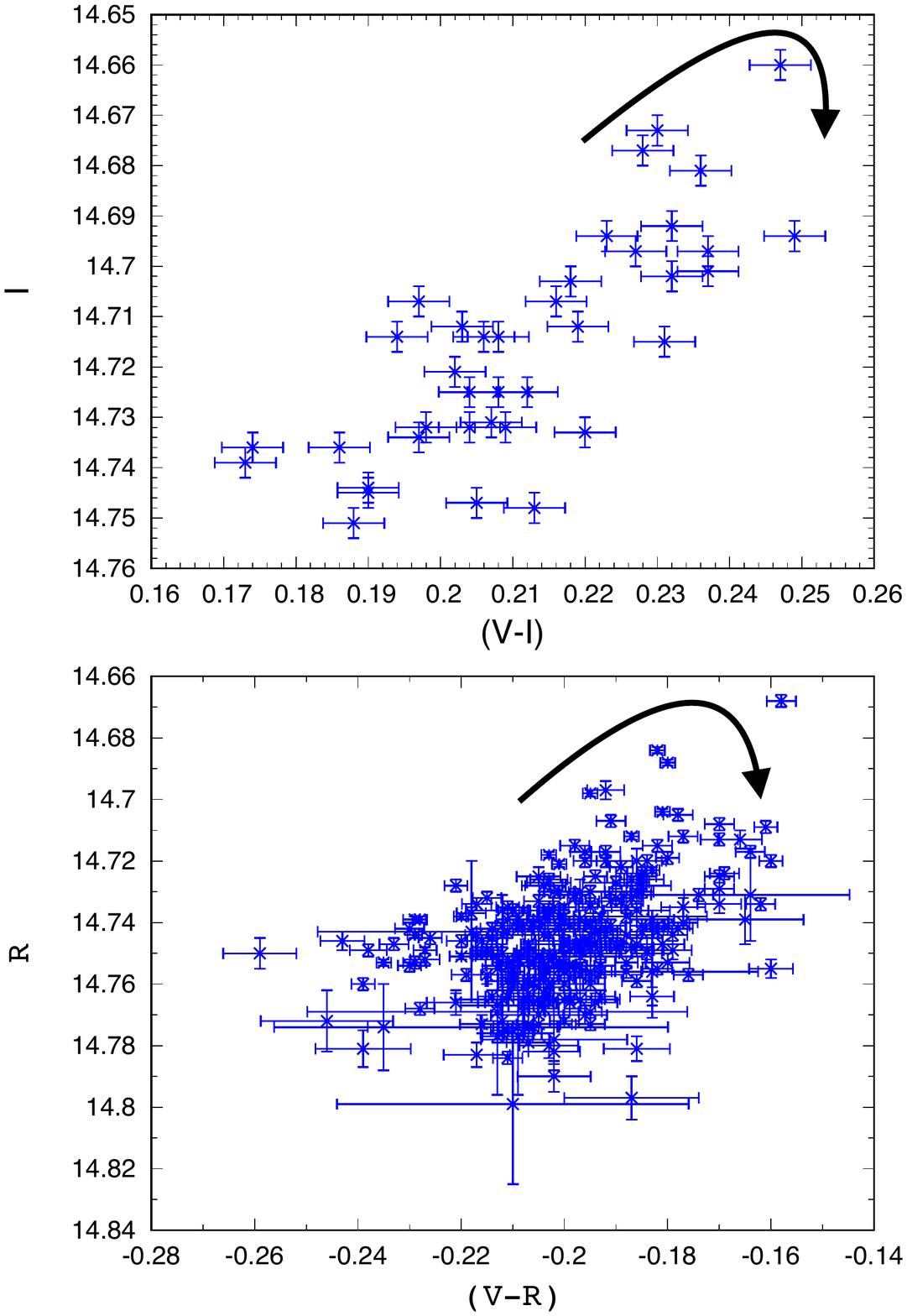}
    \caption{OGLE ($V-I$) vs $I$ (top) and MACHO $\mathcal{(V-R)}$ vs $\mathcal{R}$ (bottom) colour magnitude diagrams. A disc-less B0.5\Rom{3}$-$\Rom{5} star in the SMC has an apparent $I$-band magnitude of $\sim$15.5 and colour index of $(V-I) \sim -0.3$. The arrows indicate the direction of the evolution of the variability with time.}
    \label{fig:Colour_mag}
\end{figure}

Fig.~\ref{fig:Colour_mag} shows the historical OGLE ($V-I$)$-I$ and MACHO $\mathcal{(V-R)-R}$ colour-magnitude plots. A positive correlation between these quantities is seen in both plots, i.e. as the brightness increases the system reddens. This sort of relationship between the colour index and the magnitude is attributed to the geometry of the disc, where a relatively low inclination angle of the disc relative to the observer (i.e. orientations that are close to face-on) results in exposure to a large surface of the disc. It follows that as the disc grows in size the brightness and red continuum increase \citep{1983HvaOB...7...55H,2011MNRAS.413.1600R,2015A&A...574A..33R}. While we do not have coverage of a full cycle, we see the variability of the colour-magnitude proceed in a roughly clockwise direction. This pattern of behaviour has been seen before in Be stars and is attributed to cycles of disc growth and dissipation, as well as changes in the red and blue opacity \citep{2006A&A...456.1027D}. A disc-less star of the same spectral type as the massive companion in SXP 91.1 (B0.5\Rom{3}$-$\Rom{5}; \citealt{2008MNRAS.388.1198M}) results in an apparent $I$-band magnitude of $\sim$15.5 and colour index $(V-I) \sim -0.3$ \citep{Straizys1981,1993AcA....43..209W}, which places it off the scale of Fig.~\ref{fig:Colour_mag}.This suggests that the circumstellar disc is always present in SXP 91.1 for the duration of the observations reported here.

\subsection{Optical spectroscopy}
\begin{figure}
	\includegraphics[width=\columnwidth]{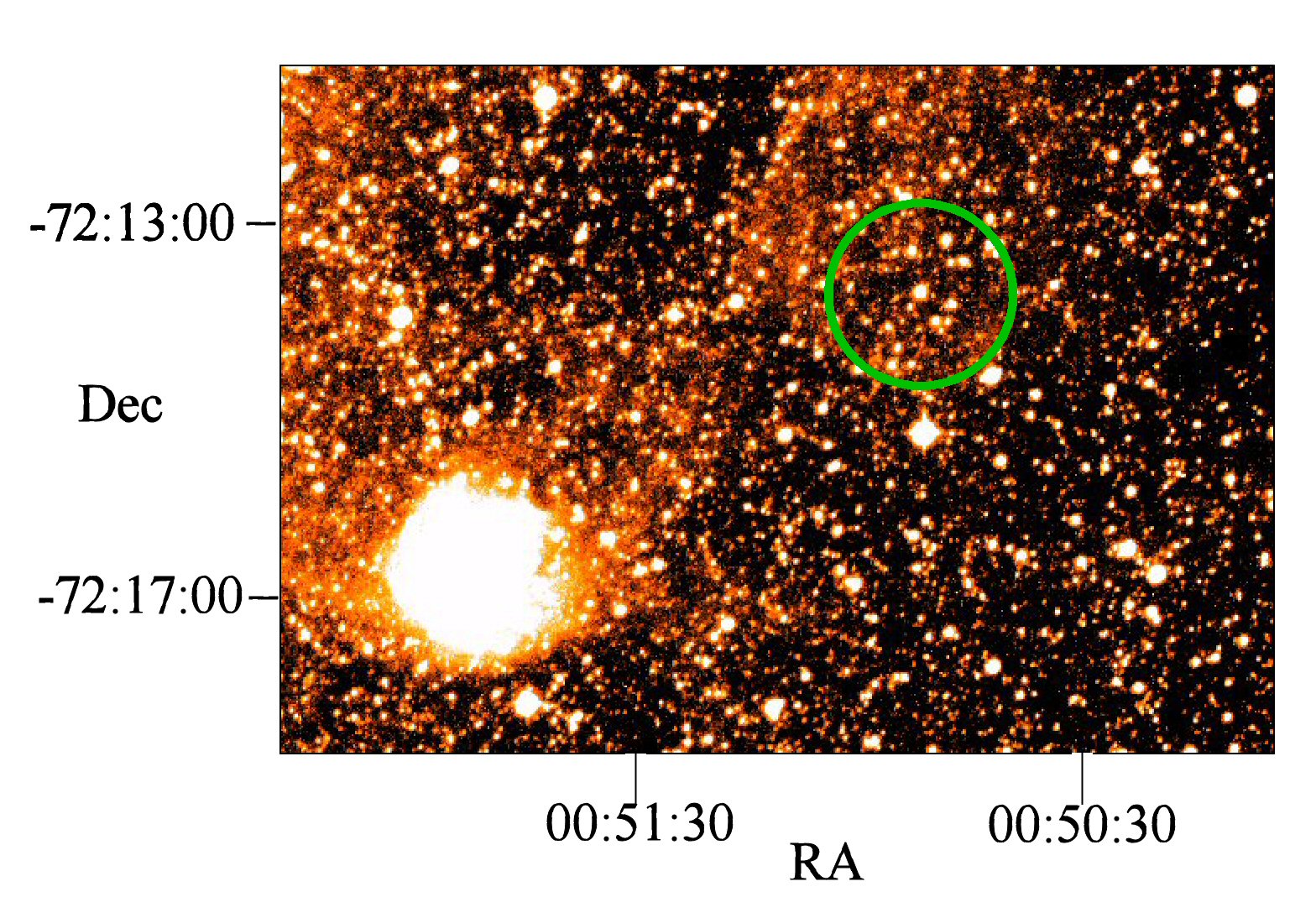}
    \caption{H$\alpha$ image of the region around SXP 91.1 from the UK Schmidt survey \citep{1998PASA...15...28P}. The source SXP 91.1 has a position of RA 00:50:55.8 Dec -72:13:38 and lies at the centre of the green circle which has a radius of 1 arcmin. }
    \label{fig:ha}
\end{figure}

\begin{figure}
	\includegraphics[width=\columnwidth]{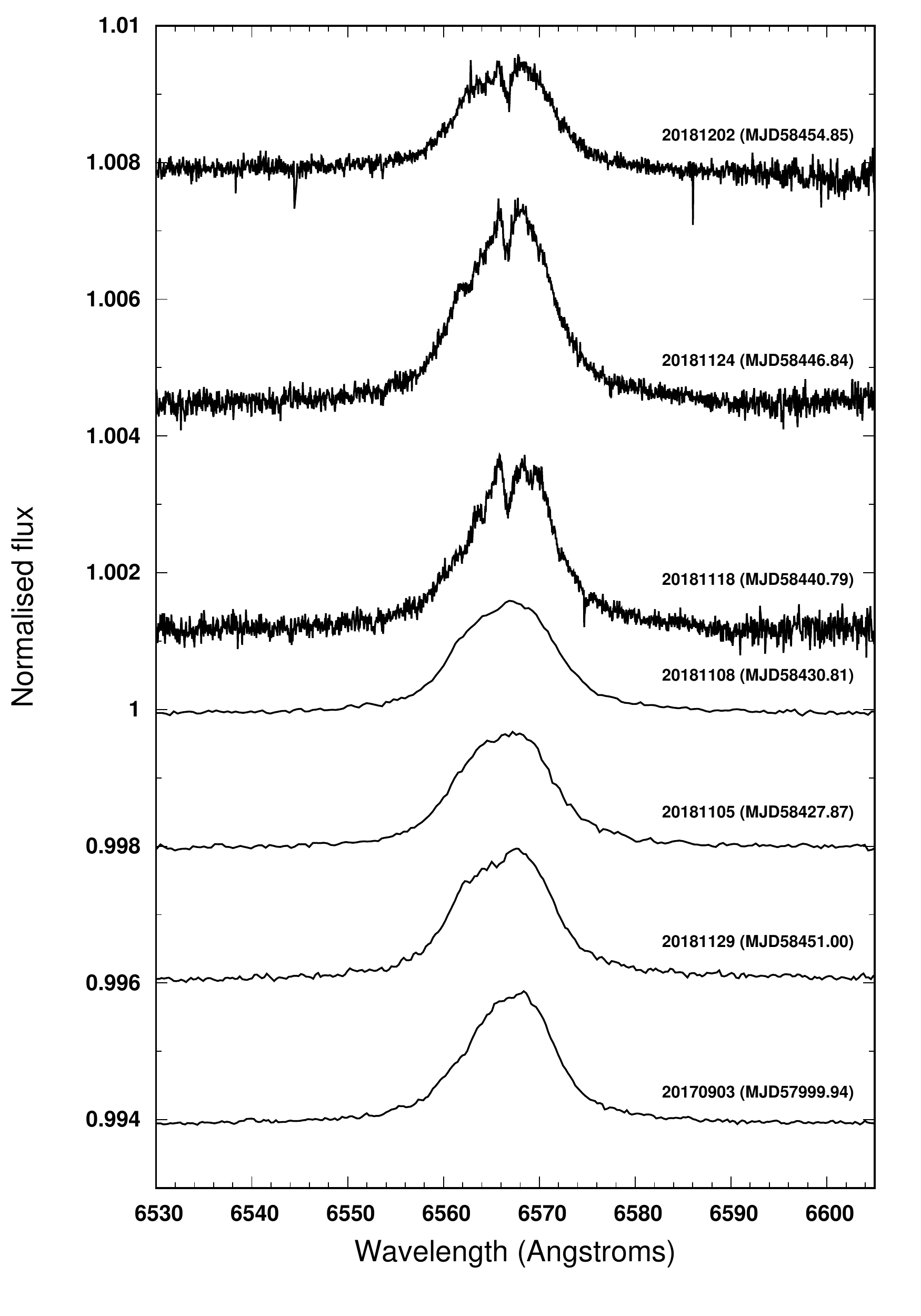}
    \caption{H$\alpha$ line profiles obtained using SALT during the current outburst. The top three were obtained using HRS and the rest are from RSS observations. The bottom spectrum is from the earliest observation, taken approximately three months before the current outburst, demonstrating the long-term disc stability.}
    \label{fig:Halpha_profiles}
\end{figure}

The SALT spectra collected for SXP 91.1 covers the H$\alpha$ line region which is seen to be in emission in all our spectra. The line profiles exhibit a double-peaked shape from the spectra obtained with the HRS and single-peaked shape in the RSS spectra. While a central depression in the HRS spectra is present due to the Keplerian distribution of the matter in the disc viewed at a non-zero inclination angle, its depth may be exaggerated due to an over-subtraction of the background emission as a result of the sky fibre being placed in the surrounding region of extended H-alpha emission (see Fig.~\ref{fig:ha}) at a distance of 1 arcmin from the target.
To demonstrate this, we re-binned the HRS spectra to reduce the resolution to match that of the RSS spectra where we see the central depression persist from the background over-subtraction.
Examples of the H$\alpha$ emission line are shown in Fig.~\ref{fig:Halpha_profiles}. The transition between different H$\alpha$ line shapes has been seen in other systems (e.g. \hbox{A~0535+262}, \citealt{2013PASJ...65...83M}) and has been explained to be caused by a precessing warped disc, resulting in the viewing angle varying periodically. The timescales for these changes, however, are much longer ($\geq 1$~year; \citealt{2013PASJ...65...83M, 2007A&A...462.1081R}) than the those seen in the spectra shown in this work for SXP 91.1 ($\leq$~10 days). The single-peaked profiles seen in the RSS spectra here are therefore likely to be due to the resolution of the RSS not being sufficient to resolve the two peaks. \\

Fig.~\ref{fig:ha} shows the extent of the H$\alpha$ emission in the region surrounding SXP 91.1. For the HRS spectra the background fibre was positioned 1 arcmin to the east of the target (the leftmost edge of the green circle in the figure). There is evidence of low level H$\alpha$ emission at this position, so the  H$\alpha$ equivalent width (EW) measurements from HRS may be a slight underestimate due to over-subtraction of the background. To account for this, the EW measurements from RSS and HRS observations obtained closest in time were used, where the difference between them is used to scale up the all the other HRS EWs. 

Fig.~\ref{fig:OGLE_EW_Scubed} shows the evolution of the EW, Swift counts and the OGLE $I$ band measurements during the current outburst. As seen in the figure, the EW evolves synchronously with the X-ray measurements. Using the EW as a proxy to the size of the disc, this suggests that the extent of the disc increases until the NS accretes and disrupts the disc matter. The measured EW values are given in Table~\ref{tab:EW_measurements}.

\begin{table}
	\centering
	\caption{H$\alpha$ equivalent width measurements of SXP 91.1}
	\label{tab:EW_measurements}
    \setlength\tabcolsep{2pt}
	\begin{tabular}{ccc} 
		\hline\hline
		MJD & EW (\AA) & Grating  \\
		\hline
57999.9405671 & 27.354 $\pm$  0.854 & PG1800 \\
58103.8093402 & 26.054 $\pm$  1.328 & PG1800 \\
58298.1372800 & 22.710 $\pm$   0.302 & PG1800 \\
58330.0717708 &  24.384 $\pm$  0.611 & PG1800 \\
58427.8736800 & 25.518   $\pm$ 0.577 & PG1800 \\
58430.8127310 & 25.608  $\pm$  0.376 & PG1800 \\
58440.7900000 & 26.498 $\pm$  0.791 & LR\\
58446.8433680 & 24.926 $\pm$  0.723 & LR\\
58451.8318866 & 24.135 $\pm$  0.612 & PG1800 \\
58454.8530210 & 23.846 $\pm$  0.522 & LR\\

		\hline
	\end{tabular}
\end{table}

\subsection{Pulse period analysis}
\begin{figure}
	\includegraphics[width=\columnwidth]{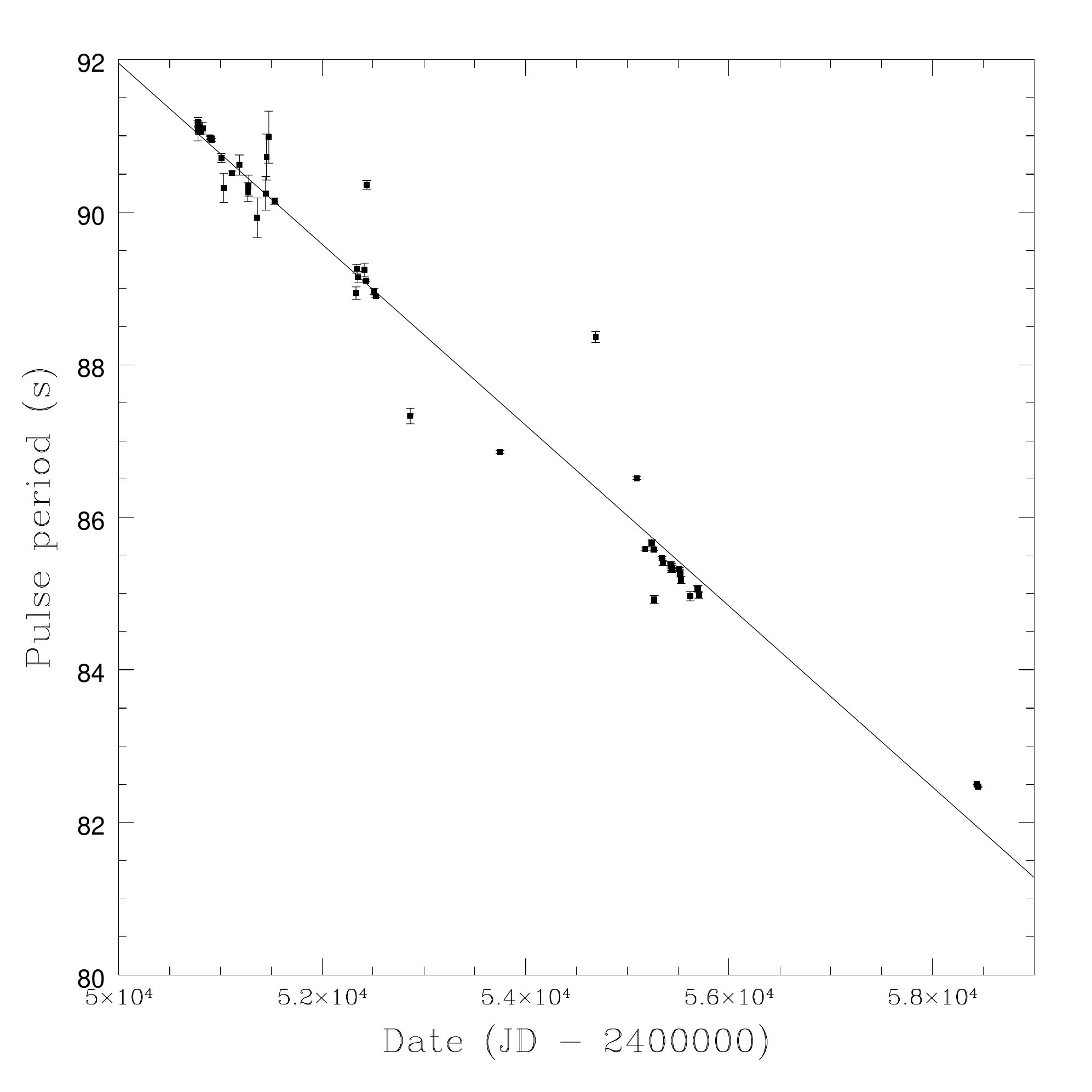}
    \caption{The pulse period history of SXP 91.1. All data points before MJD 56000 come from RXTE. There are 3 data points after this date and they are from NuSTAR. The straight line shows the weighted best linear fit to the data. See text for further details.} 
    \label{fig:pulse}
\end{figure}

The pulse period history of SXP 91.1 is shown in Fig.~\ref{fig:pulse}. All the data points shown before MJD 56000 come from observations by RXTE \citep{2013MNRAS.433...23T}. The recent data points shown come from NuSTAR observations. A summary of all the new pulse periods detected is given in Table~\ref{tab:period_measurements} which includes one detection by Swift/XRT which is not shown in the figure as the uncertainty in the pulse period is much higher than for NuSTAR.

\begin{figure}
	\includegraphics[width=1\columnwidth,angle=0]{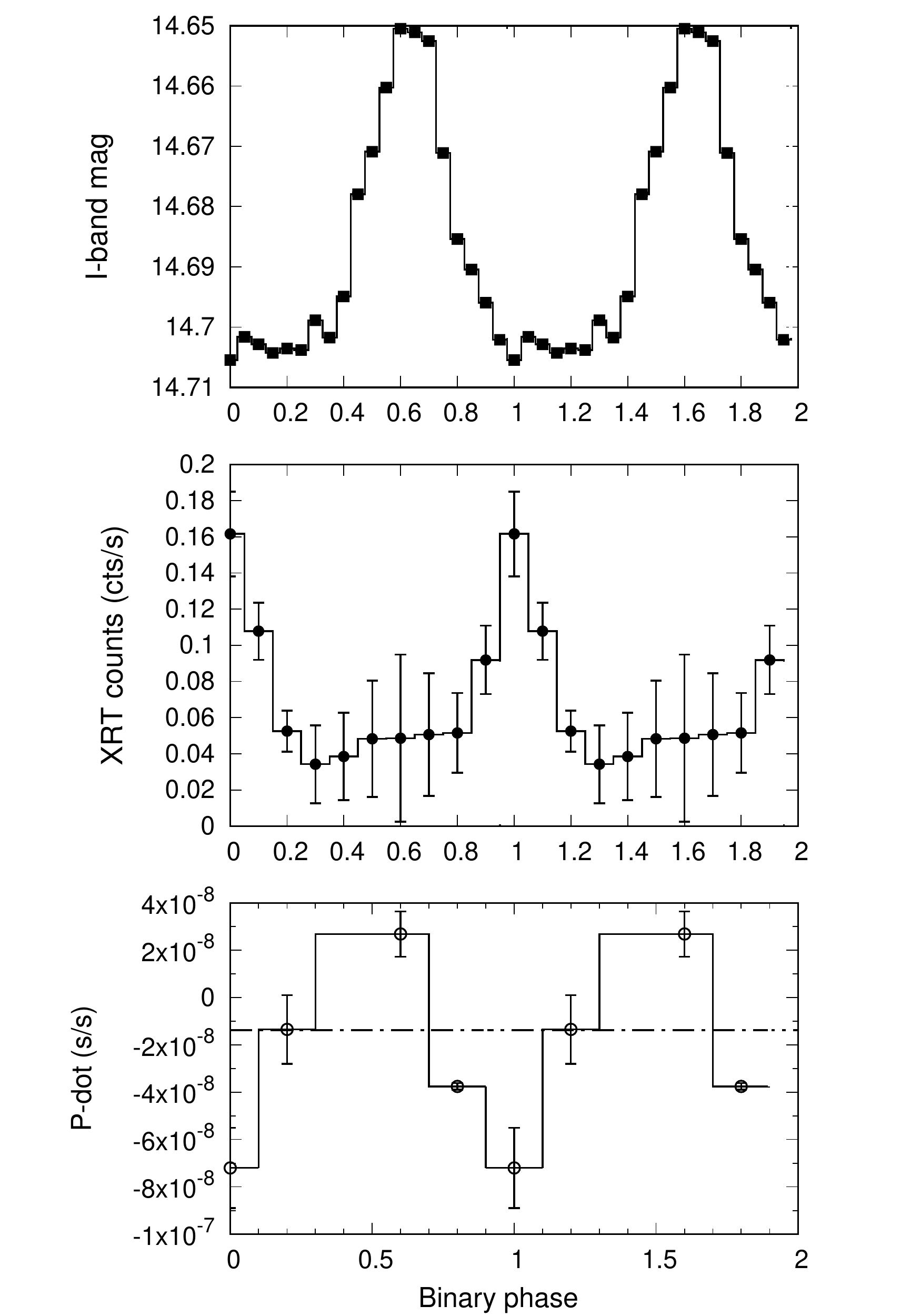}
    \caption{All OGLE (upper panel) and Swift (middle panel) data folded at the binary period. Binary phase is arbitrary but the same for both datasets. The folded $\dot{P}$ measurements weighted in phase bins of 0.2 (lower panel). The dashed horizontal line shows the mean long-term $\dot{P}$ of $1.39 \times 10^{-8}$~s.s$^{-1}$. The bottom panel shows the \textit{Swift} flux folded at the same period and phase as the upper panel.} 
    \label{fig:all_hist}
\end{figure}

A very clear and steady trend in the pulse period is shown extending over 20 years. A simple linear fit to all the data reveals a period decrease of $0.438$\,s.\,year$^{-1}$ or $1.39 \times 10^{-8}$\,s.\,s$^{-1}$. Looking at just the 3 NuSTAR points, we get a somewhat higher $\dot{P}$ value of $3.67 \times 10^{-8}$\,s.\,s$^{-1}$. An increase of some kind over the global average is probably what one would expect during the actual time of mass accretion on to the NS. This is seen in the lower panel of Fig.~\ref{fig:all_hist}, where the measured $\dot{P}$ values from the long-term pulse period are folded on the orbital period.
The $\dot{P}$ values that fall within the same phase bin intervals of 0.2 were averaged, weighted by their error bars, to produce the lower panel of Fig.~\ref{fig:all_hist}. Note that since we do not have a precise binary solution for this system we cannot say exactly where periastron occurs. Thus the binary phase shown in this figure is arbitrary. We folded the data using T$_0 = $MJD49372.5878 (the epoch of earliest observation reported in this paper from MACHO) and the orbital period of 88.37d. For the phase bin $0.3-0.5$ we do not have $\dot{P}$ measurements.
Fig.~\ref{fig:all_hist} confirms, perhaps unsurprisingly, that the NS spins up as a direct result of the mass accretion rate (seen as seen in the increase in X-ray flux).

\section{Discussion}
\subsection{Long term behaviour patterns}
\label{sec:longterm}
We can see from Fig.~\ref{fig:OGLE_multiplot} that SXP 91.1 has exhibited extremely similar optical modulation patterns for many years. If all the historical data for both OGLE (2001-2018) and S-CUBE (2016-2018) are folded at the binary period of 88.37d then we can see the average outburst profiles for both data sets - see Fig.~\ref{fig:all_hist} where we used the same T$_0$ as above.

It is clear from this figure that the relative phase offset in the X-ray and optical outbursts have, on average, persisted for many years. The \textit{I}-band outburst profile is probably a measure of the duration of the circumstellar disk distortion since most of the \textit{I}-band flux produced here and lasts, on average, for just over half the binary cycle. Whereas the X-ray outburst profile, almost certainly indicative of when material actually accretes on to the NS, is shorter, lasting just over a quarter of the binary cycle.

The relatively long duration of the $I$-band outburst suggests that the system is in a roughly co-planar and of low eccentricity since the NS would be the vicinity of the circumstellar disk, close enough to distort it, for a large fraction of the orbit. This is supported by the fact that the average X-ray profile (Fig.~\ref{fig:all_hist}) never reaches zero counts even away from the peak of the outburst, suggesting that there is low level accretion occurring far from periastron passage, which is possible for an orbit of low eccentricity.

\subsection{Pulse period history}

There are, perhaps, two surprising things about the spin period history. Firstly, the amazing consistency of the spin period change over more than 2 decades. Secondly, it is a relatively high $\dot{P}$ for a source that never seems to undergo any major outbursts, it just keeps producing Type I outbursts year after year. Several studies of different sources reveal a wide range of values of $\dot{P}$ measurements (e.g. \citealt{2001ApJ...552..738C,2001ApJ...546..455D,2011A&A...526A...7N}). As a result, its spin up rate is on the high end of $\dot{P}$ values seen from the Small Magellanic Cloud sources (see Table 1 in \cite{2014MNRAS.437.3863K} for a list of 42 such measurements).

\begin{table}
	\centering
	\caption{Recent pulse period measurements of SXP 91.1}
	\label{tab:period_measurements}
    \setlength\tabcolsep{2pt}
	\begin{tabular}{cccc} 
		\hline\hline
		MJD & Period (s) & Observatory & Exposure (ks) \\
		\hline
58431.46 & 83.293  $\pm$ 1.41 & Swift/XRT & 1.8\\
58438.68 & 82.503 $\pm$  0.001 & NuSTAR & 74.5 \\
58447.14 & 82.468 $\pm$  0.001 & NuSTAR & 38.2\\
58450.12 & 82.465 $\pm$  0.001 & NuSTAR & 73.2\\
		\hline
	\end{tabular}
\end{table}

\section{Conclusions}

This paper presents recent and long-term historic optical and X-ray observations of the SMC BeXB SXP 91.1. The long-term optical data shows that the disc has undergone stable variability, with the disc growing and shrinking with the same amplitude over two decades. This degree of stable behaviour is unusual for such systems. The system shows well-behaved consistency between the X-ray and optical flux variability over a long-term, with the peak in the X-ray occurring slightly earlier. Using the colour information, we show that the disc is always present and is oriented at low viewing angles. The long duration of the optical photometric flux suggests that the NS distorts the disc for a large portion of its orbit, implying a relatively low orbital eccentricity. As a result of the persistent accretion, the NS in SXP 91.1 has a high $\dot{P}$ which is at least an order of magnitude larger compared to a majority (\textbf{$> 60\%$}) of the known SMC sources.

\section*{Acknowledgements}

This work is partially supported by GCRF grant ref ST/R002916/1. VAM, DAHB, LJT and IMM are supported by the South African NRF. Some of the observations reported in this paper were obtained with the Southern African Large Telescope (SALT), as part of the Large Science Programme on Transients 2018-2-LSP-001 (PI: Buckley). This work is based on the research supported by the National Research Foundation of South Africa (Grant numbers 98969 and 93405). The OGLE project has received funding from the National Science Centre,
Poland, grant MAESTRO 2014/14/A/ST9/00121 to AU. This paper utilizes public domain data obtained by the MACHO Project, jointly funded by the US Department of Energy through the University of California, Lawrence Livermore National Laboratory under contract No. W-7405-Eng-48, by the National Science Foundation through the Center for Particle Astrophysics of the University of California under cooperative agreement AST-8809616, and by the Mount Stromlo and Siding Spring Observatory, part of the Australian National University.







\bsp	
\label{lastpage}
\bibliographystyle{mnras}
\bibliography{references}

\end{document}